\documentclass{article}
\usepackage[utf8]{inputenc}
\usepackage[T1]{fontenc}
\usepackage[left=1.5in, right=1.5in, top=1.3in, bottom=1.3in]{geometry}
\usepackage{mlmodern}
\usepackage{amsmath,amssymb}
\usepackage[english]{babel}
\usepackage[square, authoryear]{natbib}
\usepackage[colorlinks,citecolor=blue]{hyperref}
\usepackage[capitalize,noabbrev,nameinlink]{cleveref}
\usepackage[margin=2\parindent]{caption}
\usepackage{mathpartir,mathtools,stmaryrd}
\usepackage{enumitem,doi,xspace,minted,tikz-cd}

\title{\textbf{Internalizing Extensions \\ in Lattices of Type Theories} \\[1ex]
  \Large{\textit{Research Qualifier}}}
\author{Jonathan Chan}
\date{21 November 2024}

\crefformat{question}{#2Question~#1#3}
\Crefformat{question}{#2Question~#1#3}

\newcommand{\ie}{\textit{i.e.}\@\xspace}
\newcommand{\eg}{\textit{e.g.}\@\xspace}
\newcommand{\apost}{\textit{a posteriori}\@\xspace}
\newcommand{\kw}[1]{\mathsf{#1}}
\newcommand{\code}[1]{\texttt{#1}}
\newcommand{\HH}{\mathsf{H}}
\newcommand{\LL}{\mathsf{L}}
\renewcommand{\SS}{\mathsf{S}}
\newcommand{\oo}{\mathsf{s}}
\newcommand{\ii}{\mathsf{p}}

\begin{document}
\maketitle

\begin{abstract}
  Many proof assistants allow the use of features and axioms
  that increase their expressive power.
  However, these extensions must be used with care,
  as some combinations are known to lead to logical inconsistencies.
  Therefore, proof assistants include mechanisms that track
  which extensions are used in a proof development or module,
  ensuring that incompatible extensions are not used simultaneously.

  Unfortunately, existing extension tracking mechanisms are external to the type system.
  This means that we cannot specify precisely which extensions a definition depends on.
  Having the ability to write more precise specifications
  means we are not picking an overapproximation of the extensions needed,
  which prevents reusing definitions in the presence of incompatible extensions.
  Furthermore, we cannot refer to definitions that use incompatible extensions
  even if they are never used in inconsistent ways.
  The reasoning principles of one extension therefore cannot be used as a metatheory
  to reason about the properties of an incompatible extension.

  In this report, I explore the use of the Dependent Calculus of Indistinguishability
  (DCOI)~\citep{dcoi} for extension tracking.
  DCOI is a dependent type system with dependency tracking,
  where terms and variables are assigned dependency levels alongside their types.
  These dependency levels form a lattice that describes
  which levels are permitted to access what.
  To instead track extensions,
  each set of extensions would correspond to a dependency level,
  and the lattice would describe how extensions are permitted to interact.
\end{abstract}

\section{Introduction}

At the core of a proof assistant founded on the Curry--Howard correspondence
is a type checker that validates a proof of a proposition---a term inhabiting
a corresponding type in a dependent type theory.
These type theories are typically based on some flavour of
Martin-L\"of Type Theory (MLTT)~\citep{mltt}
or the Calculus of Inductive Constructions (CIC)~\citep{cic}.
In practice, a proof assistant does not implement merely one type theory,
but a whole host of them, as they include language extensions
that augment or modify their reasoning power.

These type theoretic extensions consist of new typing rules,
axioms, constructs, and/or definitional equalities.
Because each extension embodies semantically distinct reasoning principles,
enabling an extension results in a separate theory altogether.
Unfortunately, some extensions are mutually incompatible and cannot be enabled simultaneously
because together they violate logical consistency.
For instance, the \emph{uniqueness of identity proofs},
which propositionally equates all proofs of the same equality,
is incompatible with \emph{univalence},
which adds provably distinct proofs of equality.
In \cref{sec:extensions},
I describe these extensions and a few more found in select proof assistants,
along with the ways some combinations are incompatible.

Proof assistants are careful to track
the usage of language extensions to rule out inconsistencies.
However, the tracking done by their type checkers is \emph{external} to the type system:
within the language itself,
one cannot assert that a definition is permitted to depend on a particular extension,
nor that it is prohibited from using an extension.
Being able to specify exactly where the need for various extensions comes from
helps us be more mindful about enabling them,
especially if doing so prevents future reuse of definitions due to incompatible extensions.
Furthermore, given two incompatible extensions,
a definition in one extension's theory may not be used at all
in the another extension's theory, not even just in a type.
This means one theory cannot be used as a metatheory
to prove properties about definitions in the other theory
if those theories are incompatible.
For instance, considering classical axioms as a language extension,
one would not be able to explore what is constructively proveable
about theories that use classical principles.

Ultimately, what is missing is a framework for describing
fine-grained control of proofs and programs across multiple type theories,
where even incompatible theories can interact in interesting ways.
This fine-grained control of terms is reminiscent of
type systems with \emph{dependency analysis}.
In such systems, terms are stratified by \emph{dependency levels},
which can be thought of intuitively as permission levels tracking where terms may be used.
Even if we do not have access to a particular level,
terms at that level can still be manipulated and reasoned about opaquely
as long as they are not inspected or evaluated.

I have worked on a dependent type system with dependency tracking,
the Dependent Calculus of Indistinguishability (DCOI)~\citep{dcoi},
along with its variant DCOI$^\omega$~\citep{dcoi-omega},
which is a logically consistent type theory.
DCOI could potentially be used as a basis for a framework
that internalizes extension tracking by stratifying
the extensions' corresponding type theories into hierarchies of dependency levels,
where compatibility between extensions maps to the intuitive notion of permission.
I describe the key properties of DCOI relevant for this application in \cref{sec:dcoi}.
I then speculate on the details of this mapping,
lay out the objectives for such a framework,
and list possible first steps towards accomplishing them in \cref{sec:lattice}.
There is much prior and related work that this project relies on and relates to,
which I divide into work I have personally contributed to (\cref{sec:prior})
and other work in this space (\cref{sec:related}).

\section{Proof assistant extensions in practice} \label{sec:extensions}

To look at extensions in practice,
let us focus on three popular proof assistants:
Rocq~\citep{coq}, Agda~\citep{agda}, and Lean~\citep{lean}.
Broadly speaking, they are all based on variants of MLTT or CIC,
and have dependent functions, type universes, and inductive types.

\subsection{Built-in features}

Each of these proof assistants includes features that extend the power of their foundations;
below are a few notable extensions,
some of which are controlled by option flags.

\paragraph{Impredicativity.}
Rocq and Lean, being based on CIC,
feature a universe $\kw{Prop}$ of propositions.
This universe is \emph{impredicative},
meaning that a universal quantification (\ie dependent function type)
$\forall (x : A) \mathpunct{.} B$ is a proposition if $B$ is a proposition,
regardless of the universe in which $A$ lives,
which may be larger than $\kw{Prop}$.
Inductive types may also be defined in $\kw{Prop}$,
which permits its constructors to have argument types
in universes larger than $\kw{Prop}$.
Such inductives are said to be \emph{large}.
An impredicative $\kw{Prop}$ is a default feature of Rocq and Lean,
not an option that can be turned off.

Impredicativity allows for self-referential propositions by quantifying over $\kw{Prop}$.
For instance, given a proof that all propositions imply their double negation,
$$\mathit{dn}: \forall (P : \kw{Prop})\mathpunct{.} P \to \neg \neg P,$$
the double negation of this proposition itself holds as well by self-application.
$$\mathit{dn} \; (\forall (P : \kw{Prop})\mathpunct{.} P \to \neg \neg P) \; \mathit{dn} : \neg \neg (\forall (P : \kw{Prop})\mathpunct{.} P \to \neg \neg P)$$
In contrast, in the predicative setting,
a quantification over a universe $\kw{Type}_0 : \kw{Type}_1$ has type $\kw{Type}_1$,
and in particular,
$$\Pi (A : \kw{Type}_0)\mathpunct{.} A \to \neg \neg A : \kw{Type}_1,$$
so an element of this type may not be applied to the type itself.

Dually to universal quantification,
we can define an existential quantification $\exists (x : A) \mathpunct{.} B$
over a type $A$ in a larger universe as a large inductive proposition,
\ie a dependent pair type in $\kw{Prop}$.
This enables us to state, for example,
the surjectivity of a function over the naturals as a predicate,
even though the naturals are not a proposition.
\begin{align*}
  \mathit{surj} &: (\kw{Nat} \to \kw{Nat}) \to \kw{Prop} \\
  \mathit{surj} &\coloneqq \lambda f \mathpunct{.}
    \forall (y : \kw{Nat}) \mathpunct{.} \exists (x : \kw{Nat}) \mathpunct{.} f \; x \equiv y
\end{align*}

\paragraph{Definitional proof irrelevance.}
A universe of propositions is said to be \emph{strict}
when the inhabitants of its propositions are definitionally equal
(\ie proof irrelevant)
and thus treated as interchangeable during type checking.
Lean's $\kw{Prop}$ is always strict, while
Rocq has a separate $\kw{SProp}$ universe of proof-irrelevant propositions,
and Agda has a predicative hierarchy $\kw{Prop}_i$ of such universes \citep{sprop}.
To use strict $\kw{Prop}$,
Rocq requires the flag \code{Allow StrictProp},
while Agda requires the option \code{\{-\# OPTIONS -{}-prop \#-\}}.

Definitional proof irrelevance is useful to avoid having to prove equalities explicitly.
Consider a relation on two naturals asserting the usual less-than relation,
along with a type of bounded naturals whose boundedness can be forgotten.
\begin{align*}
  \cdot \le \cdot &: \kw{Nat} \to \kw{Nat} \to \kw{Prop} \\
  \mathit{BNat} &: \kw{Nat} \to \kw{Type}_0 \\
  \mathit{bNat} &: \Pi (n \; m : \kw{Nat})\mathpunct{.} n \le m \to \mathit{BNat} \; m \\
  \mathit{getNat} &: \Pi (m : \kw{Nat})\mathpunct{.} \mathit{BNat} \; m \to \kw{Nat} \\
  \mathit{getNat} &\; m \; (\mathit{bNat} \; n \; m \; p) \rightsquigarrow n
\end{align*}
A desirable property of bounded naturals is that
two bounded naturals are equal if their contained naturals are equal.
\begin{align*}
  \mathit{eqBNat} &: \forall (m : \kw{Nat}) \; (b_1 \; b_2 : \kw{BNat} \; m) \mathpunct{.}
    \mathit{getNat} \; m \; b_1 \equiv \mathit{getNat} \; m \; b_2 \to b_1 \equiv b_2
\end{align*}
If we try to prove this by destructing $b_1$ and $b_2$ as
$(\mathit{bnat} \; n_1 \; m \; p_1)$ and $(\mathit{bnat} \; n_2 \; m \; p_2)$,
$\mathit{getNat} \; m \; b_1$ and $\mathit{getNat} \; m \; b_2$ reduce to $n_1$ and $n_2$.
\begin{align*}
  \mathit{eqBNat} \; m \; (\mathit{bnat} \; n_1 \; m \; p_1) \; (\mathit{bnat} \; n_2 \; m \; p_2)
    &: n_1 \equiv n_2 \to \mathit{bnat} \; n_1 \; m \; p_1 \equiv \mathit{bnat} \; n_2 \; m \; p_2
\end{align*}
While we have an equality $n_1 \equiv n_2$,
we do \emph{not} have a proof of $p_1 \equiv p_2$%
\footnote{Technically, this should be a proof of the equality
where $p_1$ has been rewritten by the equality $n_1 \equiv n_2$
so that it has the same type as $p_2$.}.
Depending on how the inequality predicate $\cdot \le \cdot$ is implemented,
it may be possible to prove propositionally that any two inequality proofs are equal.
Alternatively, if propositions are definitionally proof irrelevant,
the inequality proofs can be ignored.
Then rewriting the goal by the given equality is sufficient
for it to be proven by reflexivity.
\begin{align*}
  \mathit{eqBNat} \; m \; (\mathit{bnat} \; n_1 \; m \; p_1) \; (\mathit{bnat} \; n_2 \; m \; p_2) \; e \;
    &: \mathit{bnat} \; n_1 \; m \; p_1 \equiv \mathit{bnat} \; n_2 \; m \; p_2 \\
    &\coloneqq \kw{rewrite} \; e \; \kw{in} \; \kw{refl}
\end{align*}

\paragraph{Uniqueness of identity proofs (UIP).}
The \emph{uniqueness of identity proofs} (UIP)
asserts that inhabitants of the same propositional equality
are themselves propositionally equal.
It can be proven using \emph{Axiom K}~\citep{axiomk},
a computational eliminator for propositional equalities of type $a \equiv a$.
\begin{align*}
  &K : \forall (A : \kw{Type}) \; (a : A) \; (P : a \equiv a \to \kw{Prop}) \; (p : a \equiv a) \mathpunct{.}
    P \; \kw{refl} \to P \; p \\
  &K \; A \; a \; P \; \kw{refl} \; d \rightsquigarrow d
\end{align*}
Agda's default pattern matching behaviour,
which permits matching on an equality of $a \equiv a$ as reflexivity,
admits a proof of UIP as well as defining Axiom K.
While not inherently part of Rocq's type theory,
Axiom K is axiomatized in the standard library as \code{Logic.Eqdep.eq\_rect\_eq}.

UIP is similarly useful to avoid reasoning about equalities between equalities
when the only canonical proof of an equality is reflexivity,
especially in settings without proof irrelevance.
While UIP augments the reasoning power of the type theory,
there are types whose equality proofs are already propositionally equal.
In particular, if a type has decidable equality,
\ie $(x \equiv y) \vee \neg (x \equiv y)$ for any given $x, y$ of that type,
then its equalities are themselves equal~\citep{hedberg}.

\paragraph{Strong elimination.}
Destructing or eliminating an element of an inductive datatype
into a type in a larger universe is known as \emph{strong} or \emph{large} elimination.
That is, a term whose type is in $\kw{Type}_i$
is eliminated to return a term whose type is in $\kw{Type}_j$
for some $j > i$.
For inductive proofs in $\kw{Prop}$,
this includes eliminating into any non-proposition type.

Strong elimination is a necessary ingredient in discriminating constructors
of proof-relevant datatypes, such as the booleans.
While $\kw{true}$ and $\kw{false}$ are syntactically distinct,
proving their propositional inequivalence requires lifting the booleans
to the propositions of truthhood $\top$ and falsehood $\bot$.
\begin{align*}
  \mathit{lift} &: \kw{Bool} \to \kw{Prop} \\
  \mathit{lift} &\coloneqq \lambda b \mathpunct{.} \kw{if} \; b \; \kw{then} \; \top \; \kw{else} \; \bot
\end{align*}
This branching expression is a strong elimination because it returns a type,
or equivalently because its return type is $\kw{Prop}$, a universe.
Letting $\kw{cong} \; f$ be a proof of congruence of $f$ over an equality,
to complete the proof of $\kw{true} \equiv \kw{false} \to \bot$,
given $e : \kw{true} \equiv \kw{false}$,
the trivial proof of truthhood $\kw{tt}$ is rewritten by the lifted equality
$\kw{cong} \; \mathit{lift} \; e : \top \equiv \bot$.
\begin{align*}
  \mathit{trueNotFalse} &: \kw{true} \equiv \kw{false} \to \bot \\
  \mathit{trueNotFalse} &\coloneqq \lambda e \mathpunct{.} \kw{rewrite} \; (\kw{cong} \; \textit{lift} \; e) \; \kw{in} \; \kw{tt}
\end{align*}

\subsection{Axioms}

Rocq, Lean, and Agda all have mechanisms for defining axioms or postulates,
which are declarations of constants without definitions.
Although not all axioms are consistent,
there are many well-studied axioms commonly used in practice
that are worth considering as extensions in their own right.
The axioms described here are included in many standard libraries,
and proof developments can choose whether to import them.

\paragraph{Extensionality principles.}
Some models of type theory semantically equate things
that are not syntactically (either definitionally or propositionally) equal;
extensionality principles adds semantic equalities as propositional equalities.
Examples include \emph{function extensionality},
which equates two functions if they are pointwise equal,
and \emph{propositional extensionality},
which equates two propositions if they are bi\"implicated.
\begin{align*}
  \mathit{funext} &: \forall (A : \kw{Type}) \; (B : A \to \kw{Type}) \; (f \; g : \Pi (x : A) \mathpunct{.} B \; x) \mathpunct{.}
    (\forall (x : A) \mathpunct{.} f \; x \equiv g \; x) \to f \equiv g \\
  \mathit{propext} &: \forall (A \; B : \kw{Prop}) \mathpunct{.} (A \leftrightarrow B) \to A \equiv B
\end{align*}
These axioms are found in the standard libraries of Lean as \code{funext} and \code{propext},
and of Rocq as \code{Logic.{\allowbreak}FunctionalExtensionality.{\allowbreak}functional\_{\allowbreak}extensionality}
and \code{Logic.{\allowbreak}PropExtensionality.{\allowbreak}propositional\_{\allowbreak}extensionality}.
A notable consequence of propositional extensionality is propositional proof irrelevance.

Another example is \emph{univalence},
which asserts an equivalence between propositional equality and equivalence,
\ie given two types $A, B$,
the equivalence $(A \equiv B) \simeq (A \simeq B)$ holds.
This is the core principle underlying Homotopy Type Theory (HoTT)~\citep{hott}.
There are several ways to define equivalence;
the idea is that it captures a propositionally proof-irrelevant isomorphism.
Univalence together with proof irrelevance implies propositional extensionality,
since bi\"implicated propositions are isomorphic by irrelevance,
and univalence gives an equality from the isomorphism.
Univalence alone also implies function extensionality
by a more complex argument~\cite[Chapter 4.9]{hott}.

One application of function and propositional extensionality
is encoding a function as a relation whose functionality is proven \apost.
This is a frequent pattern in proof assistants,
as inductive relations often have better ergonomic support
than dependently-typed reasoning over functions.
For example, consider a two-place predicate over
a representation of types $\mathit{Ty}$ and terms $\mathit{Tm}$,
which has the type $\mathit{Ty} \to \mathit{Tm} \to \kw{Prop}$.
If we have trouble defining this predicate recursively due to termination issues
or inductively due to strict positivity issues,
we can instead view it as a function from $\mathit{Ty}$
to a predicate $\mathit{Tm} \to \kw{Prop}$ and try encoding it as a relation:
$$R : \mathit{Ty} \to (\mathit{Tm} \to \kw{Prop}) \to \kw{Prop}.$$
Such a relation could be a \emph{logical relation}~\citep{logrel}
used to model typed lambda calculi,
where a $\mathit{Ty}$ is interpreted as a set of $\mathit{Tm}$s.
Functionality of $R$ demonstrates that $\mathit{Ty}$s have unique interpretations.
To show that $R$ is functional, \ie
$$\forall (A : \mathit{Ty}) (P \; Q : \mathit{Tm} \to \kw{Prop}) \mathpunct{.}
  R \; A \; P \to R \; A \; Q \to P \equiv Q,$$
it suffices to show that
$\forall (a : \mathit{Tm})\mathpunct{.} P \; a \leftrightarrow Q \; a$,
since $\forall (a : \mathit{Tm})\mathpunct{.} P \; a \equiv Q \; a$
follows from propositional extensionality,
and finally $P \equiv Q$ from function extensionality.

The disadvantage of axiomatic equalities is that rewriting by them does not reduce,
which can make reasoning about terms rewritten by such equalities difficult.
There are type theories beyond MLTT and CIC that are designed
so that these principles are instead provable theorems,
such as cubical type theories~\citep{bch,cchm,afh,cartesian}
and Cubical Agda~\citep{cubical-agda} for univalence,
and observational type theory~\citep{ott,ott-now,ttobs,ccobs,cicobs}
for function and propositional extensionality.

\paragraph{Classical principles.}
There are a number of classical axioms that do not hold intuitionistically.
The most common is the principle of \emph{excluded middle} (EM),
which asserts that all propositions are either true (inhabited) or false (uninhabited).
EM is equivalent to several other principles,
including \emph{double negation elimination} (DNE),
$\forall (A : \kw{Prop})\mathpunct{.} \neg \neg A \to A$,
and \emph{Peirce's law},
$\forall (A \; B : \kw{Prop})\mathpunct{.} ((A \to B) \to A) \to A$.
More powerful axioms which imply EM include the axiom of choice
and the (in)definite description operators,
which deal with extracting a concrete piece of data
out of merely knowing that such a piece of data exists.

Because a large majority of mathematics is done classically,
many communities freely use classical axioms when mechanizing mathematics.
The axiom of excluded middle, for instance,
is declared in Rocq as \code{Logic.Classical\_Prop.classic},
and in Lean as \code{em}.
The \code{Logic} subdirectory of Rocq's standard library
contains classical axioms along with proofs about their properties.
Similarly, Lean's mathematical library mathlib~\citep{mathlib}
contains proofs that rely on classical axioms,
and tactics such as \code{tauto} automatically apply classical reasoning.
% There is work towards constructively integrating classical principles into type theory consistently,
% typically in the form of calculi with control operators~\citep{control,mu,mumu,dl,sr}.

\subsection{Extensions and inconsistencies}

One has to be careful not to choose a set of features and axioms
that render the type theory logically inconsistent and thus useless for proving.
A number of them are known to be incompatible with one another;
below are a few such combinations.

\begin{itemize}[noitemsep,topsep=0pt]
  \item Strong elimination is inconsistent for large impredicative inductives.
    \citet{strong-pair} show that impredicative dependent pairs with pair projections,
    which correspond to strong elimination, are inconsistent.
    \citet{trees} also demonstrates the inconsistency of strongly eliminating
    a large inductive type $U$ with a single constructor of type
    $\forall (X : \kw{Prop})\mathpunct{.} (X \to U) \to U$.
  \item Strong elimination is also inconsistent for inductive propositions
    when $\kw{Prop}$ is proof irrelevant.
    As seen above, strong elimination suffices to show that $\neg (\kw{true} \equiv \kw{false})$.
    If $\kw{Bool}$ is defined in a proof-irrelevant $\kw{Prop}$,
    $\kw{true} \equiv \kw{false}$ would hold by definition,
    which is a contradiction.
  \item Strong elimination is once again inconsistent for inductive propositions
    in the presence of impredicativity and classical principles such as excluded middle.
    \code{Logic.Berardi} in Rocq's standard library,
    which is a modern implementation of the construction by \cite{em-irr},
    uses excluded middle to derive propositional proof irrelevance,
    which can be used as above to derive a contradiction.
  \item UIP is inconsistent with univalence.
    Intuitively, univalence produces an equality between two types
    given an equivalence between them,
    and there are types that are equivalent in multiple, provably different ways,
    so there are equalities between them that are provably different,
    thus violating UIP.
    Concretely, $\kw{Bool}$ is equivalent to itself in two different ways,
    either by mapping booleans to themselves or to their negation,
    so there are distinct proofs of $\kw{Bool} \equiv \kw{Bool}$
    \citep[Example 3.1.9]{hott}.
\end{itemize}

\begin{figure}[ht]
\centering
\begin{tikzcd}
  & \mathcolor{red}{\varnothing} & & \\
  \cdot \equiv \cdot : \kw{SProp} \arrow[ru,dashed] & & \mathrm{UA} + \kw{SProp} \arrow[lu,dashed] & \\
  \mathrm{UIP} \arrow[u] & \kw{SProp} \arrow[ru] \arrow[lu] & \mathrm{propext} \arrow[u] & \mathrm{UA} \arrow[lu] \\
  & \kw{Prop} \arrow[u] \arrow[ru] & \mathrm{funext} \arrow[ru] & \\
  & \mathrm{base} \arrow[u] \arrow[luu] \arrow[ru] & &                       
\end{tikzcd}
\caption{A compatibility graph of theories with impredicative $\kw{Prop}$, proof irrelevance ($\kw{SProp}$),
  UIP, univalence (UA), function extensionality (funext), propositional extensionality (propext),
  and (in)compatible combinations.}
\label{fig:lattice}
\end{figure}
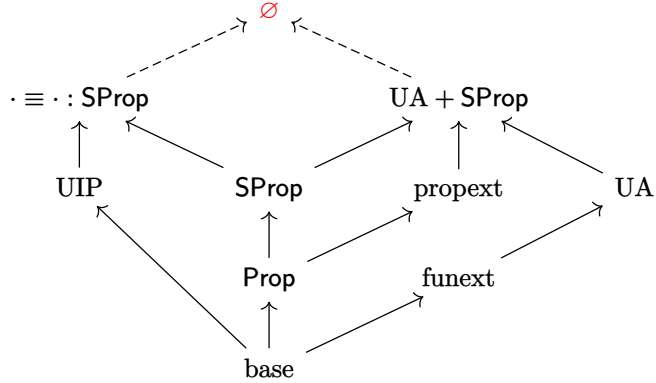

\Cref{fig:lattice} illustrates some of these relationships between extensions.
The arrows point from one theory to a strictly more expressive one;
for instance, a theory with propositional extensionality extends the equalities
of a theory containing a universe of propositions,
and a theory with univalence can derive function extensionality.
At the top of the graph, the dotted arrows indicate
the incompatibility of a theory that implies UIP with one that contains univalence:
there is no possible encompassing theory.

To prevent inconsistencies, proof assistants hide features behind option flags
or disallow them entirely.
Rocq, Lean, and Agda generally disallow strong elimination for inductive propositions
in $\kw{Prop}$ and $\kw{SProp}$.
The exceptions are \emph{syntactic subsingletons},
which are inductives that syntactically have at most one inhabitant,
such as $\top$, $\bot$, and conjunction of propositions.
While Rocq's impredicative $\kw{Prop}$ universe is not proof irrelevant,
Rocq still forbids strong elimination even for inductives that are not large
to allow the use of classical principles.
Its compiler flag \code{-impredicative-set} enables impredicativity for its $\kw{Set}$ universe
while still allowing strong elimination of small impredicative inductives,
but this flag is not well supported.%
\footnote{\url{https://github.com/coq/coq/issues/9458}}
In Agda, the \code{-{}-with-K} flag enables Axiom K
and the \code{-{}-without-K} flag disables it,
while the features enabled by Cubical Agda using the \code{-{}-cubical} flag
prove univalence as a theorem.
There is also a \code{-{}-safe} flag which disallows,
among other combinations,
having both \code{-{}-with-K} and \code{-{}-cubical}.

Proof assistants provide some limited ability in tracking the usage of features and axioms.
In Rocq, the axioms and unsafe flags used by a definition
can be listed using the command \code{Print Assumptions},
while in Lean, the axioms can be listed using \code{\#print axioms}.
In Agda, along with checking for inconsistent option combinations,
\code{-{}-safe} ensures the absence of any \code{postulate}s.
Because option flags can be enabled at module-level granularity,
Agda also has notion of \emph{(co)infective} flags:
an infective flag used in one module
must be used by all modules that depend on that module,
while a coinfective flag used in one module
may only depend on modules that also use that flag.

While users of these proof assistants have been getting along fine
with these compiler tools for tracking features and axioms for the past decade or more,
there is plenty of room of improvement;
I have identified three shortcomings these systems.
\begin{itemize}
  \item There is no way of asserting against an extension;
    that is, there is no general mechanism to tell the type checker to fail
    if a particular definition uses some feature or axiom.
    Such an assertion would guarantee that definition safe to be used by others
    that use an incompatible extension.
    For instance, ensuring that a proof does not use Axiom K
    means that it may be used by another proof that does use univalence.
    In this specific case, Agda does have the \code{-{}-without-K} flag for this purpose,
    but few options have a corresponding anti-option.
    Meanwhile, Rocq and Lean's axiom printing mechanism
    does not modify type checking behaviour.
  \item The scope of feature flags is too coarse.
    They range from project-level compiler flags
    down to module-level option flags,
    but even this level of granularity is not the appropriate one:
    modules are intended for organizing definitions by semantic content,
    rather than by the collection of features they happen to all use.
    This prevents reuse of a definition in one module inside another module
    if they happen to have incompatible features,
    even if that particular definition does not depend on any features at all.
  \item Completely disallowing an incompatible extension is unnecessarily restrictive.
    For example, when both \code{-{}-safe} and \code{-{}-with-K} are enabled in Agda,
    definitions from modules with \code{-{}-cubical} cannot be mentioned anywhere.
    However, it should be acceptable to use a theory with UIP
    to state and prove properties of univalence,
    while never eliminating an equality derived from univalence.
    To generalize, definitions from an incompatible extension
    should be permitted in the type of a term but restricted in the term itself,
    so that we can talk \emph{about} an extension without \emph{using} it.
\end{itemize}

An ideal system for enabling and disabling features and axioms, then,
should track which ones are and aren't used, at least at the definition level,
and distinguish between mentioning and using them.
This suggests that a system for dependency tracking would be a good starting point.
More precisely, the Dependent Calculus of Indistinguishability (DCOI)~\citep{dcoi},
a type system that incorporates dependency tracking and dependent types,
could be a suitable framework for tracking the usage of extensions within type theories.

\section{A primer on DCOI} \label{sec:dcoi}

DCOI is a Pure Type System (PTS)~\citep{pts} augmented with dependency tracking~\cite{dcc}.
DCOI$^\omega$~\citep{dcoi-omega} is a logically consistent instantiation
of DCOI's PTS rules and axioms with a predicative universe hierarchy,
making it suitable as a foundation for theorem proving.
In a typing judgement, dependency tracking appears as annotations
both on the variables in the context,
to indicate how they may be used by the term being typed,
and alongside the type of the term,
to indicate how the term itself may be used.
As an example, consider the following typing judgement for a constant function.
$$A :^\HH \kw{Type} \vdash \lambda x^\LL \; y^\HH \mathpunct{.} x :^\LL
  A^\LL \to A^\HH \to A$$
The concrete levels used here are low ($\LL$) and high ($\HH$) where $\LL < \HH$.
In the context of information flow,
these levels correspond to low- and high-security computations
where low-security computations may not inspect the values of high-security ones.
They can also be thought of in terms of computational irrelevance,
where something marked as computationally irrelevant ($\HH$)
must not play a part in the execution of relevant programs ($\LL$),
and may even be erased away after compilation.

This constant function at low,
which returns its first argument $x$ and ignores its second argument $y$,
must therefore mark $x$ as low to return it.
Marking $y$ as high guarantees that the function may not return it.
While the body of the low function cannot return a high argument,
its \emph{type} can depend on a high term,
demonstrated by the high-annotated type $A$ in the context,
which is used in the type of the function.
The intuition is that $A$ does not play a part in the run-time execution of the constant function,
but is otherwise permitted to participate in compile-time type checking.

Other terms we may wish to mark as high are proofs
to guarantee that a compiler can erase them away after type checking.
For instance, a length function on a computationally relevant list would return a relevant natural;
in contrast, a proof that no relevant natural is strictly less than zero would be marked as irrelevant.
\begin{align*}
  \mathit{length} &:^\LL \Pi (A :^\HH \kw{Type}) \; (l :^\LL \kw{List} \; A) \mathpunct{.} \kw{Nat} \\
  \mathit{zeroNotGt} &:^\HH \forall (n :^\LL \kw{Nat}) \; (p :^\HH n < 0) \mathpunct{.} \bot
\end{align*}

Though these examples only use two levels $\LL$ and $\HH$,
any instance of DCOI is parametrized over some meet-semilattice,
from which its dependency levels are drawn.
Rather than tracking relevance or information flow,
the goal is to track extensions
using a lattice that associates dependency levels to sets of added extensions.
Before discussing this mapping in \cref{sec:lattice},
I summarize briefly the key aspects and properties of DCOI
that highlight how dependency tracking interacts with terms and typing.

\paragraph{Relative relevance.}
The intuition of computational relevance and irrelevance
is not fixed to the low and high levels,
but is a relative concept between any two ordered dependency levels.
Suppose there is a super-high level $\SS$ such that $\LL < \HH < \SS$.
Then just as a low term may not meaningfully use a high term,
a high term also may not meaningfully use a super-high term.
The following typing judgement demonstrates an application of these three levels.
$$P :^\HH \kw{Nat}^\SS \to \kw{Prop}, n :^\SS \kw{Nat} \vdash \lambda p^\LL \mathpunct{.} p :^\LL
  (P \; n^\SS)^\LL \to P \; (n + 1)^\SS$$
In the context, $n$ is a super-high natural,
and $P$ is a high predicate which takes as argument a super-high natural.
Once again, the term being typed is a low function,
while higher terms are involved in its type.
Although the function is an identity function,
its domain and codomain types are syntactically different applications of $P$.
This judgement holds due to \emph{indistinguishability}.

\paragraph{Indistinguishability.}
In general, if $\ell_0 < \ell_1$, then at observer level $\ell_0$,
the function application $f \; x^{\ell_1}$ must be equal
to $f \; y^{\ell_1}$ regardless of what $x$ and $y$ are.
We say that they are \emph{indistinguishable} at level $\ell_0$,
written as $\cdot =^{\ell_0} \cdot$.
Below are the two different rules for indistinguishability of a function application
depending on the observer level $\ell_0$ and the argument level $\ell_1$.
\begin{mathpar}
  \infer[I-app-indist]{
    \Gamma \vdash f =^{\ell_0} g \and \ell_0 < \ell_1
  }{
    \Gamma \vdash f \; a^{\ell_1} =^{\ell_0} g \; b^{\ell_1}
  }
  \and
  \infer[I-app-dist]{
    \Gamma \vdash f =^{\ell_0} g \and
    \Gamma \vdash a =^{\ell_1} b \and
    \ell_0 \not< \ell_1
  }{
    \Gamma \vdash f \; a^{\ell_1} =^{\ell_0} g \; b^{\ell_1}
  }
\end{mathpar}
Indistinguishability plays the r\^ole of definitional equality
when checking whether two types are the same.
This is why the above example type checks,
since $P \; n^\SS =^\HH P \; (n + 1)^\SS$ holds.
Similarly, letting $k$ be the the low-level constant function above,
$k \; x^\LL \; y^\HH =^\LL k \; x^\LL \; z^\HH$ holds,
which expresses the idea that $k$ is truly constant in its second argument.

DCOI internalizes indistinguishability by indexing
its propositional equality type with an observer level,
reflected by the rule for reflexivity below.
In particular, the propositional equality
$k \; x^\LL \; y^\HH \equiv^\LL k \; x^\LL \; z^\HH$
is provable by reflexivity since the two sides are already indistinguishable
at low, the observer level of the equality.
\begin{mathpar}
  \infer[]{
    \Gamma \vdash a =^{\ell_0} b
  }{
    \Gamma \vdash \kw{refl} :^{\ell} a \equiv^{\ell_0} b
  }
\end{mathpar}

\paragraph{Elimination of higher falsehoods.}
The principle that lower-level terms may not meaningfully depend
on higher-level terms means that
destructors that produce lower-level terms may not destruct higher-level terms.
This holds even if the term being destructed contains no inner information
(such as $\top$ or an equality proof),
since reducing the destruction on a constructor requires knowing
whether the term being destructed is a constructor at all.

The sole exception is the eliminator for $\bot$,
since it has no constructors, so there is no information to reveal.
It is well typed at any level independent of its target $b$ or its type $A$.
\begin{mathpar}
  \infer[]{
    \Gamma \vdash b :^{\ell_0} \bot \and
    \Gamma \vdash A :^{\ell_1} \kw{Type}_i
  }{
    \Gamma \vdash \kw{absurd} \; b :^{\ell} A
  }
\end{mathpar}
The computational interpretation of having a proof of $\bot$ to eliminate
is that we have reached an impossible dead branch,
so what we do with it never matters since it never executes.
For example, we can define a function that extracts the head of a nonempty list
by using $\kw{absurd}$ to handle the impossible empty list case.
\begin{align*}
  \mathit{head} &:^\LL \Pi (A :^\HH \kw{Type}) (l :^\LL \kw{List} \; A) \; (p :^\HH 0 < \mathit{length} \; l^\LL) \mathpunct{.} A \\
  \mathit{head} & \; A \; \kw{nil} \; p \coloneqq \kw{absurd} \; (\kw{zeroNotGt} \; 0^\LL) \\
  \mathit{head} & \; A \; (\kw{cons} \; x \; xs) \; p \coloneqq x
\end{align*}

\paragraph{Subsumption and downgrading.}
While lower-level terms cannot inspect higher-level terms,
higher-level terms can inspect lower-level terms.
A lower-level term can also be raised to a higher level by \emph{subsumption}:
if a term is well typed at level $\ell_0$,
then it is also well typed with the same type at a higher level $\ell_1 \ge \ell_0$.
This admissible rule is given below on the left.
\begin{mathpar}
  \infer[subsumption]{
    \Gamma \vdash a :^{\ell_0} A \and \ell_0 \le \ell_1
  }{
    \Gamma \vdash a :^{\ell_1} A
  }
  \and
  \infer[downgrade]{
    \Gamma \vdash a \equiv^{\ell_1} b \and \ell_0 \le \ell_1
  }{
    \Gamma \vdash a \equiv^{\ell_0} b
  }
\end{mathpar}
However, if two terms are indistinguishable by some observer level $\ell_1$,
then they are indistinguishable by a \emph{lower} observer level $\ell_0$ by \emph{downgrading},
given above on the right.
From a security flow perspective, the higher the observer level,
the more secure values may be observed,
and the more things are distinguishable,
since more secure values need to be compared instead of being ignored.
Going down an observer level means more things are being hidden away,
so more values appear to be indistinguishable from one another.

\section{Lattices of type theories} \label{sec:lattice}

The key premise of using DCOI for extension tracking
is associating dependency levels with various type theories,
creating a lattice whose points are the sets of extensions
enabled in each theory, ordered by subset.
We begin with a bottom dependency level for the base type theory
corresponding to the empty set of extensions.
For each additional construct corresponding to a feature or axiom,
we add a new dependency level above the theory it extends.
For instance, there could be a K eliminator that type checks at a level for UIP,
or a built-in excluded middle axiom that type checks at a level for classical reasoning.
\begin{align*}
  \kw{K} \; P \; p \; d :^{\kw{uip}} P \; p &&
  \kw{em} \; A :^{\kw{cl}} A \vee \neg A
\end{align*}
\vspace{-1.5\baselineskip}

Because level annotations are part of contexts and typing judgements,
well-typedness of a particular definition also specifies exactly where it is safe to be used,
guaranteeing that it never exploits an extension without permission.
A definition that can be typed at the bottom level
would be safe to be used at all levels by subsumption,
and guaranteed to never employ, say, classical reasoning.
Indistinguishability reflects this guarantee, as it asserts the property that
uses of terms from higher forbidden theories can only be trivial,
such as ignoring them or passing them around uninspected.

As dependency levels form a meet-semilattice,
any two theories must have a meet (\ie intersection),
which corresponds to only the constructs that they both have in common,
and which are therefore safe to use in either theory.
If the join (\ie union) of two theories exist,
then the constructs introduced in either one can be used at the joined level.
Crucially, not all joins exist; a UIP level cannot be joined with a univalence level,
since their co\"existence is contradictory.
The shape of the lattice depends on the compatibilities between theories,
as well as the implication order of extensions,
since one theory that encompasses the consequences of another
can be placed above that other theory.
The compatibility graph in \cref{fig:lattice}
is an example of a concrete lattice of theories,
where the arrows point towards the greater theory
and indicate the direction in which definitions can be raised.

The property that the type of a term can itself
be well typed at a level independent of the term's
permits proving propositions about an incompatible theory
without causing an inconsistency.
For example, we can assert the computational behaviour of the K eliminator
even in a theory with univalence.
\begin{align*}
  \kw{refl} :^{\kw{ua}} \kw{K} \; P \; \kw{refl} \; d \equiv^{\kw{uip}} d
\end{align*}
\vspace{-1.5\baselineskip}

Following the rules of DCOI,
each individual theory must be logically consistent.
If an inconsistency exists at any theory,
by the elimination of higher falsehoods,
the inconsistency propagates to all lower theories, including the bottom theory.
Then by subsumption, the inconsistency at the bottom theory
can be raised to propagate to all higher theories,
and the entire lattice would be inconsistent.
This means that if eliminating falsehoods works exactly as in DCOI,
any theory that features nontermination would not be permitted.

As the goal is to exclude incompatible extensions from a proof assistant,
disallowing logically inconsistent theories is a desirable trait.
Nevertheless, there may be a few ways to modify falsehood elimination to permit them.
One way is to disallow eliminating falsehoods to lower levels,
allowing it only to the same level.
Another is to take ideas from works from the Trellys project,
such as $\lambda^\theta$~\citep{lambda-theta} and Sep$^3$~\citep{sep3},
and impose a value or termination restriction on falsehoods being eliminated.
If the falsehood in an inconsistent theory is nonterminating or not a value,
then it cannot be eliminated at all,
preventing its propagating to lower theories.

One catch is that a theory whose extension is a new definitional equality
(\ie a new rule for indistinguishability) cannot be contained within its level.
Even if that equality is defined for a particular observer level,
it will hold for all lower observer levels by downgrading,
and the extension will be available to all lower theories.
This effect cannot be mitigated using restrictive premises,
as violating downgrading violates many other desirable properties,
including transitivity of definitional equality~\citep{dcoi-omega}.

Consequently, adding strictness to an existing $\kw{Prop}$ universe
rather than adding an entirely separate $\kw{SProp}$ universe is not be possible,
as the type checker can lift two proofs at a non--proof-irrelevant level
up to the proof-irrelevant level and equate them there.
One solution is to invert the conventional order
and place non-strict $\kw{Prop}$ \emph{above} strict $\kw{Prop}$,
so that \emph{disabling} strictness is an extension.
The inversion slightly complicates the lattice in \cref{fig:lattice},
as the level with propositional UIP via the K eliminator
would be above the level with propositional equality in non-strict $\kw{Prop}$,
which in turn would be above the level with equality in strict $\kw{Prop}$
and thus with definitional UIP.
Such a lattice has the unusual property that the univalent level
would be joinable with the non-strict $\kw{Prop}$ level,
but not with the UIP or strict $\kw{Prop}$ levels above and below it.

\subsection{Objectives}

This project should answer the following questions:

\begin{enumerate}
  \item \label[question]{item:extensions}
    What kinds of extensions would fit within this framework?
    Some broad classifications of extensions might be
    ones that add new type universes (\eg $\kw{SProp}$),
    ones that expand the rules for existing constructs
    (\eg impredicativity, strong elimination),
    ones that add new computational constructs with reduction rules
    (\eg Axiom K),
    and ones that add new axiomatic constructs without reduction rules
    (\eg function and propositional extensionality, excluded middle).
  \item \label[question]{item:applications}
    What are useful applications of being able to freely refer to other theories?
    The previous example that referred to the K eliminator, while true,
    is stating a trivial fact that is already provable within the UIP theory.
    Are there meaningful theorems about one theory that cannot be proven in that theory,
    but can be proven in a different yet potentially incompatible theory?
  \item \label[question]{item:model}
    How would a particular lattice of theories be modelled
    to show logical consistency?
    Ideally, the technique used to model a particular lattice
    should be broadly applicable and sufficiently extensible
    to a different lattice without redoing all the work,
    so that adding more extensions remains sustainable.
\end{enumerate}

To answer these questions, the project would be divided into two portions.
The first is an implementation of a type checker for a specific lattice of type theories.
The lattice should contain a sufficiently diverse set of labels and their orders.
\Cref{fig:lattice} is a good place to start,
as it contains theories in different classifications with different interactions.

To evaluate the viability of such a type checker,
a standard library would be implemented to exercise all levels of the lattice.
The standard libraries of Rocq\footnote{\url{https://coq.inria.fr/distrib/current/stdlib/}},
Agda\footnote{\url{https://agda.github.io/agda-stdlib/master/}},
and Lean\footnote{\url{https://leanprover-community.github.io/mathlib4_docs/}}
are good sources for inspiration,
as many of their files use the features and axioms mentioned in \cref{sec:extensions}.
An implementation would also serve to verify which extensions are indeed invalid
by demonstrating the inconsistencies they yield.

Additionally, writing and checking large proofs would be feasible,
which helps with exploring more complex applications.
One example is computing \emph{semisimplicial types},
which is an open problem in HoTT.
Two-level type theory (2LTT)~\citep{2ls,2ltt} is one solution
that contains separate theories with univalence and with UIP,
and Agda implements 2LTT as the \code{-{}-two-level} extension.
We can use libraries\footnote{\eg \url{https://github.com/UnivalencePrinciple/2LTT-Agda}}
for 2LTT in Agda as inspiration for testing whether this project's implementation
of orthogonal univalence and UIP extensions can handle the same proof load.
\Cref{sec:multi} discusses some more details about 2LTT.

The useability of the implementation would inform the design of the system,
such as level inference and level polymorphism.
Annotating definitions and arguments with every single extension it uses
is an unreasonable burden on a proof assistant user.
The prototype implementation accompanying DCOI~\citep{dcoi-artifact}
therefore has rudimentary level inference, defaulting to a minimum level.
However, the prototype's lattice is the usual order of the naturals,
and its examples typically use no more than two levels.
An implementation with a lattice containing incompatible extensions
and applications that involve more than two theories
would pinpoint what is required from level inference in practice
for a more sophisticated inference algorithm.

Rewriting libraries in this implementation is also an exercise
in determining where code duplication occurs
and whether level polymorphism would help eliminate it.
Although subsumption allows lifting definitions vertically,
so to speak, from lower theories to higher ones,
it does not allow transporting definitions horizontally
from one theory to a different, incompatible theory.
In addition, while a function can quantify over terms at specific levels,
it cannot quantify over \emph{all} levels,
nor over levels that satisfy some ordering constraint.
Having a library of examples in the implementation
would reveal whether these are real concerns to be addressed
and what kind of level polymorphism could address them.

% With an implementation, useability concerns can be explored,
% such as level inference.
% Annotating definitions and arguments with every single extension it uses
% is an unreasonable burden on a practical proof assistant user,
% and it may be possible to infer the annotations
% either based on the syntactic constructs used
% or on what set of features are required for successful type checking.

The second portion is a formalized and ideally mechanized proof of consistency.
Because consistency is a semantic property and depends on the strength
of the metatheory used to model the type theory,
the formalization should model a lattice with (at least at first)
only one level above the base theory, the simplest nontrivial lattice.
The focus would be on how to combine two different models of type theory,
not on accommodating as many as possible from the outset.

A sensible starting point would be taking the mechanization of DCOI$^\omega$~\citep{dcoi-omega},
which proves consistency and normalization of what would be the base theory in the lattice,
and picking a reasonable feature to extend it with.
UIP and function extensionality are good candidates,
as they are expected to already hold in the semantic model.
In this case, the challenge is to prove that the base theory does \emph{not}
prove UIP or function extensionality to demonstrate that
other incompatible extensions could be added to the base theory.

Moreover, the semantic model is a syntactic logical relation
indexed by well-founded universe levels,
which may limit its extensibility;
it cannot be straightforwardly extended to accommodate impredicativity,
nor to accommodate typed definitional equality.
If the PTS rules and axioms of DCOI are instead instantiated to
a single impredicative universe with no universe hierarchy or strong elimination,
we can adapt the logical relation for the Calculus of Constructions by \citet{cc}
to prove consistency and normalization.

A possible alternative is to use \emph{syntactic} modelling~\citep{syntactic},
which would involve a type-preserving translation into another type theory
whose consistency is well established,
guaranteeing consistency of the original system.
While there exist syntactic models of other type theories \citep{sprop,ghosts}
with notions of irrelevance, which is one application of indistinguishability,
a syntactic model of dependency tracking with dependent types is unexplored.

% The process of accomplishing these two portions of the project
% should answer \cref{item:properties},
% either by the implementation revealing unexpected examples that can or cannot be type checked,
% or by metatheoretical properties that hold based on the modelling technique chosen.
% Only once these properties are revealed will we know what further work can be done,
% from augmenting the implementation closer to a practical proof assistant,
% to proving more complex theorems like normalization and decidability of type checking,
% or proving consistency for a larger lattice.

\section{Prior work} \label{sec:prior}

This project builds on prior work on DCOI~\citep{dcoi} and DCOI$^\omega$~\citep{dcoi-omega},
on both of which I am second author.
For the former paper,
I implemented a prototype type checker for DCOI augmented with inductive types
by extending the minimal dependent type checker \texttt{pi-forall}~\citep{pi-forall},
and I wrote examples using the type checker and motivating examples for DCOI.
I also proved a few of the lemmas in the mechanization.
For the latter paper,
I wrote about half of the prose, mostly for the earlier sections,
and proved a few of the lemmas as well.
As part of an investigation toward incorporating a relational model for DCOI,
I mechanized a PER model for MLTT based on the logical relation
used to prove consistency of DCOI$^\omega$,
but ultimately the gap between MLTT and DCOI could not be bridged,
so this work does not appear in the final paper.

Outside of DCOI, I have worked on Stratified Type Theory (StraTT)~\citep{stratt},
which annotates typing judgements similarly to dependency tracking,
but the annotations are universe levels.
Instead of stratifying universes into a hierarchy,
typing judgements themselves are stratified,
and there is a single universe whose type is itself.
To ensure consistency in the presence of this \emph{type-in-type} rule,
dependent functions may only quantify over types at strictly lower levels,
which enforces predicativity.
Although StraTT is not a dependency tracking system in the same way DCOI is,
it demonstrates that there may be multiple ways
to retain usage information that enforces consistency.
Even if the particular setup for DCOI turns out not to be suitable for this project,
it may be reasonable to instead explore a more StraTT-like structure.

\section{Related work} \label{sec:related}

\subsection{Multi-system frameworks} \label{sec:multi}

\paragraph{Two-level type theory.}
The most similar work to extension tracking is two-level type theory (2LTT)~\citep{2ls,2ltt}.
It consists of an inner homotopical type theory with univalence
and an outer intensional type theory with UIP,
along with a conversion operation $\mathop{\uparrow} \cdot$ from the inner theory to the outer.
The inner and outer type theories have independent type formers,
including separate inner (path) equality types $\cdot \equiv^\ii \cdot$
and outer (strict) equality types $\cdot \equiv^\oo \cdot$.
Converting an inner equality does \emph{not} yield the outer equality type;
otherwise, univalence on inner equalities could be converted to univalence on outer equalities,
which would contradict UIP of the outer equality.

These inner and outer levels are different from dependency levels in DCOI,
where all levels share the same type formers,
and a lower equality can be raised to a higher equality by subsumption.
In particular, if a lattice of type theories includes one that supports UIP,
then that level will prove that proofs of the same equality at \emph{all} levels are equal.
Meanwhile, in 2LTT, UIP only holds for proofs the outer equality
and not for converted proofs of the inner equality.

The conversion operator can be thought of as an explicit subsumption,
and how it interacts with the inner and outer equalities
is similar to how DCOI's propositional equality interacts with indistinguishability
at lower and higher levels.
To demonstrate, given two inner terms $x, y$,
the implication $\mathop{\uparrow} x \equiv^\oo \mathop{\uparrow} y \to x \equiv^\ii y$
holds in 2LTT while the converse generally does not.
Similarly, in DCOI, $x \equiv^\HH y \to x \equiv^\LL y$ holds by downgrading
while the converse also generally does not.

\paragraph{The Trellys project.}
Instead of combining multiple type theories,
the Trellys project focussed on combining dependently-typed logical reasoning
with (potentially nonterminating) functional programming.
The main works within the project are $\lambda^\theta$~\citep{lambda-theta},
which classifies typing judgements of a single language into logical and programmatic fragments;
Zombie~\citep{zombie}, an implementation that closely follows $\lambda^\theta$;
Sep$^3$~\citep{sep3}, which syntactically separates proofs from programs;
and Nax~\citep{nax}, which augments dependent types with Mendler-style recursion schemes.

Of these three, DCOI is closest to $\lambda^\theta$,
whose logical and programmatic classifications are similar to DCOI's dependency levels.
Because the logical fragment is subsumed within the programmatic fragment,
the additional features found in the programmatic fragment can be thought of
as an extension of the logical one.
Notable features of the extension include isorecursive types and unrestricted recursion,
allowing for nonterminating programs.
Normalization of the logical fragment is ensured by
only allowing boxed programs to be applied to its functions,
only allowing unboxing of values,
and restricting reduction to call by value.
While the boxing mechanism is similar to level annotations on function domains in DCOI,
the value restrictions are specific to handling the presence of potential divergence.

\Citet{proofs-programs} extends the work done for $\lambda^\theta$
by looking at a number of languages with logical and programmatic fragments,
beginning with a simply typed calculus with classifications (also named $\lambda^\theta$),
extending it with types dependent on terms in LF$^\theta$,
then further extending that with terms dependent on types (\ie type polymorphism)
and types dependent on types (\ie type-level computation) in PCC$^\theta$.
The proofs of normalization for these calculi use partially step-indexed logical relations,
where stepping only occurs in the programmatic fragment.
However, the proof technique does not scale up to handle large elimination or a universe hierarchy,
both of which are present in Zombie.

\paragraph{System DE.}
Similarly to PCC$^\theta$, System DE~\citep{system-de}
is a dependent type system with logical and programmatic fragments,
where the logical fragment is again normalizing.
It extends System DC~\citep{system-dc}
and is designed to be suitable as a core calculus
of the Glasgow Haskell Compiler~\citep{ghc};
as such, it deals with explicit coercion proofs.
In contrast to PCC$^\theta$,
System DE does not permit using a programmatic value as a logical value,
so its logical relation proof does not require partial step-indexing
to interpret the programmatic fragment.
Consequently, System DE is able to include
the type-in-type rule and large elimination in the programmatic fragment.

\subsection{Other proof assistants}

\Cref{sec:extensions} broadly covers a number of optional features
and common axioms in Rocq, Lean, and Agda.
There are many other proof assistants of varying relevance not discussed above.

Idris 2~\citep{idris2} is a dependently typed programming language with partiality.
Definitions can be marked as \code{total}, \code{covering}, or \code{partial};
setting aside type-in-type, totality ensures consistency,
covering ensures type safety while allowing divergence,
and partiality does not ensure either.
Because partiality subsumes covering subsumes totality,
these modifiers can also be thought of as members of a lattice.
Similarly to PCC$^\theta$, Idris 2 is also call by value,
and the partiality modifier is designed so that
diverging terms can appear in types and be reasoned about
while not being reduced during type checking.
Although Idris 2 is based on Quantitative Type Theory~\citep{qtt},
it has no complete formal description that describes all of its core features,
especially as it is a rapidly evolving language.

F$^\star$~\citep{fstar} is a proof assistant with dependency tracking for different effects.
Its dependency levels include a \code{Tot} level for total programs
and a \code{Dv} level above it for potentially diverging programs.
In contrast to PCC$^\theta$,
the total fragment of F$^\star$ may not refer to the diverging fragment,
so the proof of weak normalization of the total fragment
involves a logical relation that does not consider levels above \code{Tot}.
All effects in F$^\star$ are implemented as indexed monads,
which get compiled away to the new core calculus TotalF$^\star$~\citep{total-fstar};
divergence aside, effects do not extend the internal type system.

\section{Conclusion}

In this report, I have described a number of common extensions
to proof assistants that alter their core type theory when enabled.
These proof assistants have mechanisms for tracking extensions
and ensuring that incompatible extensions may not be used together.
However, extension tracking is external to the type system,
and internalization opens up opportunities
for greater precision and expressivity in specifying extension usage.
Such an internalization of extension tracking is reminiscent of dependency analysis,
and in particular of the Dependent Calculus of Indistinguishability (DCOI),
which tracks usage of terms at different dependency levels.
Although primary applications of DCOI are information flow and irrelevance,
I use it as inspiration for a framework that tracks extensions in the same way,
incorporating multiple type theories in one system.
Much work lies ahead to discover the practical expressivity of such a system,
as well as its logical consistency,
for its viability as a core for a proof assistant with extensions.

\clearpage
\noindent
\bibliography{main}
\end{document}